\documentclass[american, aps,prl, twocolumn, showpacs,reprint]{revtex4} 
\usepackage[unicode=true,pdfusetitle, bookmarks=true,bookmarksnumbered=false,bookmarksopen=false, breaklinks=false,pdfborder={0 0 0},backref=false,colorlinks=false] {hyperref}
\hypersetup{ colorlinks,linkcolor=myurlcolor,citecolor=myurlcolor,urlcolor=myurlcolor}
\usepackage{braket,colortbl,cleveref,amsthm,amsmath,amssymb,txfonts}
\definecolor{myurlcolor}{rgb}{0.0,0.0,0.9}
\usepackage{graphics,graphicx}
\usepackage{color}
\usepackage{graphicx}
\usepackage{graphicx}  
\usepackage{booktabs}
\usepackage[font=small,labelfont=bf,
   justification= centerlast,
   format=hang]{caption}
\usepackage{subfigure}
\usepackage{tabularx}
\usepackage{amsmath}
\DeclareMathOperator{\tr}{Tr}
\usepackage{graphicx}
\usepackage[utf8x]{inputenc}
\usepackage{color}
\usepackage{amsmath}
\usepackage{braket}
\usepackage{latexsym}
\usepackage{amssymb}
\usepackage{amsthm}
\usepackage{bm}
\usepackage{graphics,epstopdf}
\usepackage{color}\usepackage{amsmath}
\usepackage{enumitem}
\usepackage{fmtcount}
\usepackage{booktabs}

\usepackage{epsfig}

\theoremstyle{plain}

\def\bea{\begin{eqnarray}}
\def\eea{\end{eqnarray}}
\def\ba{\begin{array}}
\def\ea{\end{array}}

\def\ket{\rangle}
\def\bra{\langle}
\def\beq{\begin{equation}}
\def\eeq{\end{equation}}

\begin{document}

\title{Quantifying coherence with quantum addition }

\author{Chiranjib Mukhopadhyay}
\author{Arun Kumar Pati}
\author{Sk Sazim}
\address{Quantum Information and Computation Group, Harish Chandra Research Institute, Homi Bhabha National Institute, Allahabad 211019, India}

\begin{abstract} 
\noindent 
 
Quantum addition channels have been recently introduced in the context of deriving entropic power inequalities for finite dimensional quantum systems. We prove a reverse entropy power equality which can be used to analytically prove an inequality conjectured recently for arbitrary dimension and arbitrary addition weight. We show that the relative entropic difference between the output of such a quantum additon channel and the corresponding classical mixture quantitatively captures the amount of coherence present in a quantum system. This new coherence measure admits an upper bound in terms of the relative entropy of coherence and is utilized  to formulate a state-dependent uncertainty relation for two observables. Our results may provide deep insights to the origin of quantum coherence for mixed states that truly come from the discrepancy between quantum addition and the classical mixture.

 
\end{abstract}

\maketitle
\emph{I\lowercase{ntroduction}}- The superposition principle lies at the heart of quantum theory. There are two ways of exploring superposition as a keystone for quantum technologies. The first is to consider a systematic resource theoretic approach, developed in the last few years as the \emph{resource theory of quantum coherence}\citep{aberg,baumgratz,uttament,winter, roc1, roc2,max,deba,mcms,colloquium,lin1,lin2,lin3}. Several generalizations \citep{theurer,noc,regula} and links to other quantum resources like entanglement\citep{uttament}, nonclassicality\citep{vedral,mile,killoran}, magic \citep{chiru9} or superradiance\citep{kwek} have been successfully established through this approach. The other approach is to analyze  quantum superposition in a physical context, seeking to pinpoint how the linear superposition principle makes quantum physics qualitatively different from classical physics.

For pure states, it is natural that quantum coherence arises from superposition. However, for mixed states, there is no such notion of coherent superposition of mixed states. On the other hand, one can have classical mixture of two or more mixed states. Therefore, it is  \emph{a priori} unclear how quantum coherence can arise from the superposition principle for mixed states. Here, we show that for mixed states, the quantum coherence actually arises from the notion of quantum addition, which allows us to include an additional contribution arising from the non-commutativity of the mixed components over and above the classical mixture. In this paper, we prove that the quantum coherence naturally comes from the discrepancy between quantum and classical addition. This shows for the first time how quantum coherence arises for mixed states from an analogus notion of superposition, namely, the quantum addition.

 Shannon originally proposed \citep{shannon} an \emph{entropic power inequality (EPI)} for classical continuous random variables, where the addition of two random variables was taken in the sense of convolution. This inequality was later proved by Stam \citep{stam} by the use of the Fisher information and the de Bruijn identity. A quantum version of the EPI for CV systems was proposed recently by Koenig and Smith \citep{ks} and proved along similar lines. In this version, the addition is in the sense of addition of modes of the optical field, which may be experimentally implemented via beamsplitters. The finite dimensional, i.e., qudit analogue to the EPI was finally proved \citep{doa,doa2} recently, where the authors introduced a quantum addition operation, which is, in some sense, the qudit analogue of the operation of a beamsplitter. 

In this letter, after  a brief recapitulation of the qudit addition channel and the resource theory of quantum coherence, we first demonstrate reverse entropic relations for the qudit addition channel, which we use to prove a conjecture furnished in Ref. \citep{chinese} for arbitrary choice of  parameters. Subsequently, we move on to quantifying coherence in a quantum system through a quantity which arises naturally as a deficit term from the reverse entropic relation derived. We show by demonstrating the requisite monotonicity properties, that this quantity, which we term as the \emph{coherence of quantum addition (CQA)} is indeed a coherence quantifier. We establish an upper bound to the CQA in terms of the well-known relative entropy of coherence \citep{baumgratz}. We finally move on to establishing a quantum uncertainty relation with a state-dependent lower bound utilizing the CQA before concluding with a discussion on the impact of the present work and the scope for future works.

\emph{P\lowercase{reliminaries}}- In this section, we briefly recapitulate the basic framework on which the present work is based. We begin by reviewing the qudit addition channel as introduced in Ref. \citep{doa}. Subsequently we discuss the basic features of the resource theory of coherence useful in this context.

\emph{Quantum addition and entropic power inequalities-} Shannon originally conjectured the entropy power inequality for continuous random variables $X,Y$ in terms of the differential entropy $H$ in the following form 

\beq e^{H(X * Y)} \geq e^{H(X)} + e^{H(Y)}\eeq

\noindent where * is the convolution operation. Various proofs of this inequality, beginning with the original proof by Stam \citep{stam} have come up over the years \citep{epi1,epi2,epi3}.  A quantum generalization of the above inequality was obtained for CV systems considering the beamsplitter merging operation as the analogue of convolution. Dutta, Ozols and Audenart  \citep{doa} have recently found the EPI for finite dimensional quantum systems through the  qudit addition channel \citep{lloyd}. Essentially, this channel can be realized as a unitary evolution in  $\mathbb{C}^d \otimes \mathbb{C}^d$ followed by tracing out the ancilla qudit. Let $\rho \in \mathcal{B}(\mathbb{C}^d), \sigma \in \mathcal{B}(\mathbb{C}^d)$  be  two qudit states. Then the quantum addition channel is a completely positive trace preserving (CPTP) map \beq \tr_2\left[  U_{\alpha} (\rho \otimes \sigma) U_{\alpha}^{\dagger}\right] = \rho \boxplus_{\alpha} \sigma = \alpha \rho + (1-\alpha) \sigma - i \sqrt{\alpha (1-\alpha)} [\rho,\sigma], \eeq where the unitary $U_{\alpha}$ is the \emph{partial swap} channel, i.e. $U_{\alpha} = \sqrt{\alpha}\  \mathbb{I} + i \sqrt{1-\alpha} \ S$, where $S$ is the two qudit swap gate $ \sum_{i,j=1}^{d} |i\ket\bra j|\otimes |j\ket\bra i|$. The Kraus operators corresponding to this map are expressible in the form \beq K_{n} = \sqrt{\lambda} \mathbb{I} \otimes \bra n| + i \sqrt{1-\lambda} \bra n| \otimes \mathbb{I}. \eeq It has been proved in \citep{doa} that for any concave function $f$ which depends solely on the spectrum of a state $\rho$, the following relation holds true for $\alpha \in [0,1]$
\beq 
f(\rho \boxplus_{\alpha} \sigma) \geq \alpha f(\rho) + (1-\alpha) f(\sigma).
\eeq

\noindent If $f$ is chosen as the exponential of the von Neumann entropy, we obtain the qudit analog to the classic EPI. For another suitable choice of $f$, the resulting inequality is a qudit analog of the entropic photon number inequality \citep{doa}. 
\begin{figure}
\includegraphics[width = 0.35 \textwidth, keepaspectratio]{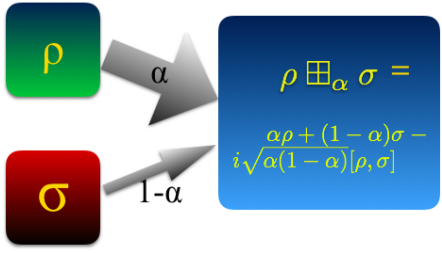}
\caption{A schematic diagram of the quantum addition channel}
\end{figure}

\emph{Quantum coherence-} The resource theory of quantum coherence seeks to formally quantify the amount of superposition possessed by a quantum state with respect to some fixed basis. Like any convex resource theory, it has a set of free states called \emph{incoherent states} which are diagonal states in the given basis, and a set of free operations, which are called \emph{incoherent operations (IO)}. Although other families of incoherent operations like \emph{maximally incoherent operations (MIO)} \citep{chitambar}, \emph{strictly incoherent operations (SIO)} \citep{vedral}, \emph{physical incoherent operations (PIO)} \citep{chitambar,chitambar2} or \emph{genuinely incoherent operations (GIO)}\citep{genuine} have been put forward in the literature, we shall restrict ourselves to incoherent operations (IO) only. Since PIO,SIO or GIO are restricted subsets of incoherent operations, the proof of monotonicity under IO automatically imply monotonicity under these operations.

If a quantity $C(\rho)$ is to be considered as a quantifier of quantum coherence for a state $\rho$ under IO, it must satisfy first three of the following four properties, with the fourth one, i.e., convexity being a desired but not an essential feature.

\begin{enumerate}
\item \textit{Faithfulness}- $C(\rho) = 0$ iff $\rho$ is an incoherent state, i.e., diagonal in the given basis.

\item \textit{Weak monotonicity}- Under any incoherent channel $\Lambda$, the amount of quantum coherence never increases, i.e., $C(\Lambda(\rho)) \leq C(\rho)$.

\item \textit{Strong monotonicity-} If $\lbrace K_i \rbrace$ are Kraus operators corresponding to any incoherent operation $\Lambda$ such that $\sigma_{i} = \frac{K_{i} \rho K_{i}^\dagger}{\tr[K_{i} \rho K_{i}^\dagger]} = \frac{K_{i} \rho K_{i}^\dagger}{p_i} $, then the amount of coherence should not increase under selective measurement, i.e., $\sum_i p_i C(\sigma_i) \leq C (\rho)$.

\item \textit{Convexity-} An additional desirable feature of a coherence quantifier is convexity over the states, i.e., if $\rho = \sum_{i} p_i \rho_i$, then, $C(\sum_{i} p_{i} \rho_i) \leq \sum_{i} p_{i} C(\rho_i) $.
\end{enumerate}

Proving the strong monotonicity condition in particular, can be quite challenging. Therefore the following condition has been developed \citep{tong} as an equivalent condition to the third and fourth criteria provided the first and second criteria are met. 
\beq C(p_1 \rho_1 \oplus p_2 \rho_2) = p_1 C(\rho_1) + p_2 C(\rho_2)\ \text{with  } p_1,p_2 \geq 0, p_1 + p_2 =1 \eeq
In this work, we shall use this criterion to establish the new coherence quantifier as a monotone. 

\emph{A \lowercase{Reverse Entropic Power Relation} -} The entropic power inequality puts a lower bound on the entropy of the output of the qudit addition channel. In this section, we attempt to find an upper bound to the entropy of the output of the qudit addition channel. \\

\noindent \emph{\textit{Theorem (Reverse EP equality)}- If $\rho$ and $\sigma$ are two qudit states and $\boxplus_{\alpha}$ denotes quantum addition with weight $\alpha \in (0,1)$, then the following equality holds} \beq 
S(\alpha \rho + (1-\alpha) \sigma) = S(\rho \boxplus_{\alpha} \sigma || \alpha \rho + (1-\alpha) \sigma) + S(\rho \boxplus_{\alpha} \sigma) \label{rev_epi}
\eeq

\begin{proof}
Let us first analyze the difference in von Neumann entropies between the results of classical mixture and  quantum addition of states $\rho$ and $\sigma$. 
\begin{widetext}
\begin{align}
& S(\alpha \rho + (1-\alpha) \sigma) - S(\rho \boxplus_{\alpha} \sigma) \nonumber \\
&= S(\rho \boxplus_{\alpha} \sigma + i \sqrt{\alpha (1-\alpha)} [\rho, \sigma]) - S(\rho \boxplus_{\alpha} \sigma)  \\
&= \tr\left[(\rho \boxplus_{\alpha} \sigma) \log (\rho \boxplus_{\alpha} \sigma) \right] - \tr\left[(\rho \boxplus_{\alpha} \sigma + i \sqrt{\alpha (1-\alpha)} [\rho, \sigma]) \log(\rho \boxplus_{\alpha} \sigma + i \sqrt{\alpha (1-\alpha)} [\rho, \sigma]) \right] \nonumber  \\
&=  \tr\left[(\rho \boxplus_{\alpha} \sigma) \log (\rho \boxplus_{\alpha} \sigma) \right] - \tr\left[(\rho \boxplus_{\alpha} \sigma) \log(\rho \boxplus_{\alpha} \sigma + i \sqrt{\alpha (1-\alpha)} [\rho, \sigma]) \right] - \tr\left[i \sqrt{\alpha(1-\alpha)} [\rho,\sigma] \log(\alpha \rho + (1-\alpha) \sigma) \right] \nonumber \\
&= S(\rho \boxplus_{\alpha} \sigma || \alpha \rho + (1-\alpha) \sigma) - \tr\left[i \sqrt{\alpha(1-\alpha)} [\rho,\sigma] \log(\alpha \rho + (1-\alpha) \sigma) \right] 
\end{align}
\end{widetext}
Now, let us concentrate on the second term. Since $\alpha \rho + (1-\alpha) \sigma$ is a positive semidefinite matrix, its logarithm, say $K$, is Hermitian and commutes with $\alpha \rho + (1-\alpha) \sigma$. Thus, \beq [K,\rho] = -\frac{1-\alpha}{\alpha} [K, \sigma] \label{comm}\eeq

Armed with this result, we now simplify the second term in the following way.
\begin{align}
\tr\left[[\rho,\sigma] \log(\alpha \rho + (1-\alpha) \sigma) \right]  &= \tr\left[\rho \sigma K\right] - \tr\left[\sigma \rho K \right]\nonumber\\
&= \tr[\sigma K \rho] - \tr[\sigma \rho K] \nonumber\\
&= \tr\left[\sigma [K,\rho]\right]\nonumber \\
&= -\frac{1-\alpha}{\alpha}\tr\left[ \sigma[K,\sigma] \right]\nonumber\\
&= -\frac{1-\alpha}{\alpha} \tr\left[ \sigma K \sigma - \sigma \sigma K\right]  \nonumber \\
&= 0.
\end{align}
Here we have used the cyclicity of trace as well as the commutation relation in \eqref{comm}. Thus, the second term in the remainder vanishes and the proof is complete.
\end{proof}
\noindent An immediate corollary to the theorem above is the following inequality for arbitrary $\alpha \in  [0,1]$, which was conjectured  \cite{chinese} to hold for arbitrary qudit systems when $\alpha = \frac{1}{2}$. 

\noindent \emph{\textit{Corollary-} If $\rho$ and $\sigma$ are two qudit states, then } \beq S(\rho \boxplus_{\alpha} \sigma) \leq S(\alpha \rho + (1-\alpha) \sigma). \eeq

This result follows from the relation \eqref{rev_epi} and the non-negativity of quantum relative entropy. It is easy to see that for $\alpha \in (0,1)$, this inequality is strict if $\rho$ and $\sigma$ do not commute.This we term as the reverse entropy power inequality. Thus quantum addition of two density operators results in a density operator whose entropy is always lower compared to the corresponding classical mixture. One may thus wonder, does this difference in entropy capture some quantumness ? In the rest of the paper, we confirm that this is indeed the case and this difference quantitatively captures the amount of quantum coherence present in a state. Similar reverse entropic power inequalities can be constructed from elementary entropic properties and will be considered elsewhere.

\emph{C\lowercase{oherence of quantum addition}} - We now propose the following quantity, which we name as the \emph{coherence of quantum addition (CQA)}, as a coherence quantifier for every $0 \leq \alpha \leq 1$. This is defined as
\beq
C_{\alpha} (\rho) := \min_{\sigma \in \mathcal{I}}S \left( \rho \boxplus_\alpha \sigma || \alpha \rho + (1-\alpha) \sigma \right) \ .
\eeq

Below we show that this quantity is a coherence monotone. That this quantity vanishes for incoherent $\rho$ is obvious. In order to proceed towards proving the monotonicity of $C_\alpha$ under IO channels , let us first prove the following lemma - 

\noindent \emph{\textit{Lemma-} If $\sigma$ is an incoherent state and $\Lambda$ denotes a strictly incoherent channel, then the following equality holds}
\beq 
\Lambda(\rho) \boxplus_\alpha \Lambda(\sigma) = \Lambda(\rho \boxplus_\alpha \sigma)
\eeq

\noindent\emph{Proof-} The Kraus operators $\lbrace K_i \rbrace$ corresponding to the IO $\Lambda$ are known to be doubly stochastic and thus, may be expressed in the form $ K_{i} = \sum_{k} d_{ik} |\pi_{k}\ket\bra k|$, where $|\pi_{k}\ket$ is a permutation of basis element $|k\ket$.  LHS of the equality to be proved reads as  $ = \alpha \Lambda(\rho) + (1-\alpha) \Lambda(\sigma) - i \sqrt{\alpha (1-\alpha)} [\Lambda(\rho), \Lambda(\sigma)]$.  Now, let us focus on the quantity $\Lambda(\rho) \Lambda(\sigma)$, which equals 
\begin{align}
\Lambda(\rho) \Lambda (\sigma)  &= \sum_{ij} K_{i} \rho K_{i}^{\dagger} K_{j} \sigma K_{j}^{\dagger} \nonumber \\
& =  \sum_{ij} \sum_{mnt} K_{i} \rho d_{im}^* |m\ket\bra \pi_m | d_{jn} |\pi_n\ket\bra n| \sigma d_{jt}^*|t\ket\bra \pi_{t}|  \nonumber \\
&= \sum_{ij} \sum_{mnt} K_{i} \rho d_{im}^* d_{jn} d_{jt}^* |m\ket \bra \pi_{m} | \pi_n \ket \bra n| \sigma |t \ket \bra \pi_t | \nonumber \\
&= \sum_{ij} \sum_{mt} K_{i} \rho d_{im}^* d_{jm} d_{jt}^* |m\ket\bra m| \sigma |t \ket \bra \pi_t | \nonumber \\
&= \sum_{ij} \sum_{mt} K_{i} \rho  \sigma d_{im}^* d_{jm} d_{jt}^* |m\ket\bra m| t \ket \bra \pi_t | \nonumber \\
&= \sum_{ij} \sum_{m} K_{i} \rho  \sigma d_{im}^* d_{jm} d_{jm}^*  |m \ket \bra \pi_m | \nonumber \\
&= \sum_{i} \sum_{m} K_{i} \rho \sigma d_{im}^* \left( \sum_{j} d_{jm} d_{jm}^* \right) |m\ket \bra \pi_m| \nonumber \\
&= \sum_{i} \left( K_{i} \rho \sigma \sum_{m} d_{im}^* |m\ket \bra \pi_{m}| \right)  \nonumber \\
&= \sum_{i}K_{i} \rho \sigma K_{i}^{\dagger} \nonumber \\
&= \Lambda (\rho \sigma)
\end{align}

Similarly one can prove that $\Lambda(\sigma \rho) = \Lambda(\sigma) \Lambda (\rho)$. The equality to be proved now follows straightforwardly.
 \qed

Having proved this lemma, we can immediately prove the monotonicity condition under incoherent channels. Before that, let us remark that this lemma, which shows that the qudit addition a.k.a. partial swap channel commutes with an IO, may be of independent interest. In particular, for CV settings,  the beamsplitter operates as a similar channel to the qudit addition channel described here. Thus, it may be interesting to check whether the free operations in resource theory of coherence for CV systems do also commute with the beamsplitter channel, when one of the inputs in the beamsplitter happen to be free, e.g, a quantum optical coherent state or a thermal state.

\noindent\emph{\textit{Theorem (Monotonicity under incoherent channels)} Under any incoherent channel $\Lambda$, and for any state $\rho$, the following relation holds for all $\alpha \in [0,1]$, i.e.,} \beq C_\alpha (\Lambda(\rho)) \leq C_\alpha(\rho).\eeq

\noindent\emph{Proof-} Suppose $\sigma$ is the  incoherent state which yields the requisite minimization for $C_{\alpha} (\rho)$. Now, \begin{align}
& C_{\alpha} (\Lambda (\rho))  \leq S(\Lambda(\rho) \boxplus_{\alpha} \Lambda(\sigma) || \alpha \Lambda(\rho) + (1-\alpha) \Lambda(\sigma)) \nonumber \\
& = S(\Lambda(\rho \boxplus_{\alpha} \sigma) || \Lambda (\alpha \rho + (1-\alpha) \sigma )) \nonumber \\
& \leq S(\rho \boxplus_{\alpha} \sigma || \alpha \rho + (1-\alpha) \sigma ) \nonumber \\
&= C_{\alpha}(\rho)
\end{align}
Here we have used the previous lemma as well as the fact that the quantum relative entropy is a monotone under CPTP maps. 
\qed

\noindent We now move on to proving the following property, which, in conjunction with the monotonicity property proved earlier, establishes the CQA as a full coherence monotone.
\vspace{0.1 in}

\noindent\emph{\textit{Theorem (Equality under direct sum)} For two states $\rho_1$ and $\rho_2$, and probabilities $p_1,p_2 \in [0,1]$ such that $p_1 + p_2 =1$, the following equality holds for all $\alpha \in [0,1]$} \beq C_\alpha (p_1 \rho_1 \oplus p_2 \rho_2) = p_1 C_\alpha( \rho_1) + p_2 C_\alpha ( \rho_2)\ . \eeq

\noindent \emph{Proof-} 
We shall prove this equality by first proving it as an inequality in one direction and then proving it as an inequality in the opposite direction. Suppose $\sigma_1$ and $\sigma_2$ are incoherent states yielding the required minimization for evaluation of CQA for states $\rho_1$ and $\rho_2$ respectively. Let us write the LHS as
\begin{widetext}
\begin{align*}
& C_\alpha (p_1 \rho_1 \oplus p_2 \rho_2)  \leq S((p_1 \rho_1 \oplus p_2 \rho_2) \boxplus_\alpha (p_1 \sigma_1 \oplus p_2 \sigma_2)|| \alpha (p_1 \rho_1 \oplus p_2 \rho_2) + (1-\alpha) (p_1 \sigma_1 \oplus p_2 \sigma_2))\nonumber \\
& = S(p_1 (\rho_1 \boxplus_\alpha \sigma_1) \oplus p_2 (\rho_2 \boxplus_\alpha \sigma_2) || p_1 (\alpha \rho_1 + (1-\alpha) \sigma_1) \oplus p_2 (\alpha \rho_2 + (1-\alpha) \sigma_2) ) \nonumber \\
&= p_1 S(\rho_1 \boxplus_\alpha \sigma_1|| \alpha \rho_1 + (1-\alpha) \sigma_1) + p_2 S(\rho_2 \boxplus_\alpha \sigma_2|| \alpha \rho_2 + (1-\alpha) \sigma_2) \nonumber \\
&= p_1 C(\rho_1) + p_2 C(\rho_2)
\end{align*}

\end{widetext}
Now, we have to prove this inequality in the opposite direction. Let us assume the incoherent state $\sigma$ for which the requisite minimization for $C_{\alpha} (p_1 \rho_1 \oplus p_2 \rho_2)$ is obtained, is expressible as \beq \sigma = q_1 \sigma_1 \oplus q_2 \sigma_2; \hspace{0.2 in} q_1 + q_2 = 1\eeq

\noindent Thus, $ C_\alpha (p_1 \rho_1 \oplus p_2 \rho_2)  = S((p_1 \rho_1 \oplus p_2 \rho_2) \boxplus_\alpha (q_1 \sigma_1 \oplus q_2 \sigma_2)|| \alpha (p_1 \rho_1 \oplus p_2 \rho_2) + (1-\alpha) (q_1 \sigma_1 \oplus q_2 \sigma_2)) = p_1 S(\rho_1 \boxplus_{\alpha} \frac{q_1}{p_1} \sigma_1 || \alpha \rho_1 + (1-\alpha) \frac{q_1}{p_1} \sigma_1)) + p_2 S(\rho_2 \boxplus_{\alpha} \frac{q_2}{p_2} \sigma_2 || \alpha \rho_2 + (1-\alpha) \frac{q_2}{p_2} \sigma_2)).$ Now let us assume without loss of generality that $ \frac{q_1}{p_1} \geq 1$, which automatically implies $\mu = \frac{q_2}{p_2} \leq 1$.  Let us now concentrate on the second term of the above expression. 

\begin{widetext}
\begin{align}
& p_2 S(\rho_2 \boxplus_{\alpha} \frac{q_2}{p_2} \sigma_2 || \alpha \rho_2 + (1-\alpha) \frac{q_2}{p_2} \sigma_2)) = p_2 S((\rho_2 \oplus 0) \boxplus_{\alpha} (\mu \sigma \oplus (1-\mu) \mathbb{I} || \alpha (\rho_2 \oplus 0) + (1-\alpha)( \mu \sigma \oplus (1-\mu) \mathbb{I})   ) \nonumber \\
& \geq p_2  C_{\alpha} (\rho_2 \oplus 0)  = p_2 C_{\alpha} (\rho_2)
\end{align}

\end{widetext} Here the first equality follows from the property $S(A \oplus B || C \oplus D) = S(A||C) + S(B||D) $ followed by the relative entropy. The subsequent equality of $C_{\alpha}(\rho_2 \oplus 0)$ and $C_{\alpha}(\rho_2)$ can be shown from the monotonicity condition under IO which has already been proved earlier.  Now, we have to prove that $S(\rho_1 \boxplus_{\alpha} \frac{q_1}{p_1} \sigma_1 || \alpha \rho_1 + (1-\alpha) \frac{q_1}{p_1} \sigma_1)) \geq C_\alpha(\rho_1)$. Suppose the Kraus operators corresponding to the qudit addition channel and the convex mixing channel are $\lbrace K_i \rbrace$ and $\lbrace L_i \rbrace$ respectively. Then, 
\begin{widetext}
\begin{align}
&S(\rho_1 \boxplus_{\alpha} \frac{q_1}{p_1} \sigma_1 || \alpha \rho_1 + (1-\alpha) \frac{q_1}{p_1} \sigma_1)) = S\left(\sum_{i} K_{i} (\rho_1 \otimes \frac{q_1}{p_1} \sigma_1)K_{i}^{\dagger} || \sum_{i} L_{i} (\rho_1 \otimes \frac{q_1}{p_1} \sigma_1)L_{i}^{\dagger}\right) \nonumber \\
&= \frac{q_1}{p_1} S\left( \sum_{i} K_{i} (\rho_1 \otimes \sigma_1)K_{i}^{\dagger} || \sum_{i} L_{i} (\rho_1 \otimes  \sigma_1)L_{i}^{\dagger}\right) = \frac{q_1}{p_1} S(\rho_1 \boxplus_{\alpha} \sigma_1 || \alpha \rho_1 + (1-\alpha) \sigma_1 ) \geq \frac{q_1}{p_1} C_{\alpha}(\rho_1) \geq C_{\alpha} (\rho_1)
\end{align}
\end{widetext}
\noindent Thus, we have \beq C_{\alpha}(p_1 \rho_1 \oplus p_2 \rho_2) \geq p_{1} C_{\alpha} (\rho_1) + p_{2} C_{\alpha} (\rho_2) \ . \eeq

Since we have already proved this inequality the opposite way, the equality condition follows.
 \qed
 \\
\emph{Relating the coherence of quantum addition to relative entropy of coherence-} Now that we have established the coherence of quantum addition as a full coherence monotone, the next task is to relate it to already extant measures of coherence. One of the most important measures of quantum coherence is the relative entropy of coherence. This has operational interpretation as the coherence cost in the asymptotic setting \citep{winter}, as well as the cost associated with erasing quantum coherence \citep{avijit}. Below we give an upper bound to the CQA in terms of the relative entropy of coherence.

\noindent\emph{\textit{Theorem -} coherence of quantum addition is upper bounded by a function of the relative entropy of coherence $C_r$ in the following way}
\beq 
C_{\alpha}(\rho) \leq \frac{h(\alpha)}{\sqrt{2}} \sqrt{C_{r} (\rho)} \ ,
\eeq
\noindent \emph{where $h(\alpha)$ is the binary entropy function and equals $ -\alpha \log \alpha - (1-\alpha) \log (1-\alpha)$.}

\begin{proof}
Let us start with the entropic power inequality \citep{doa} \beq S(\rho \boxplus_{\alpha} \sigma) \geq \alpha S(\rho) + (1-\alpha) S(\sigma)\eeq
Combining this inequality with \eqref{rev_epi} yields, 
\begin{widetext}
\begin{align}
& S(\rho \boxplus_{\alpha} \sigma || \alpha \rho + (1-\alpha) \sigma) = S(\alpha \rho + (1-\alpha) \sigma) - S(\rho \boxplus_{\alpha} \sigma) \nonumber \\
& \leq S(\alpha \rho + (1-\alpha) \sigma) - \alpha S(\rho) - (1-\alpha) S(\sigma) \nonumber \\
&= \alpha \tr\left[ \rho \log \rho - \rho \log(\alpha \rho + (1-\alpha) \sigma) \right] + (1-\alpha)  \tr\left[ \sigma \log \sigma - \sigma \log(\alpha \rho + (1-\alpha) \sigma) \right] \nonumber \\
&= \alpha S(\rho || \alpha \rho + (1-\alpha) \sigma) + (1-\alpha) S(\sigma || \alpha \rho + (1-\alpha) \sigma) \nonumber \\
& \leq h(\alpha) \frac{1}{2} ||\rho - \sigma||_1 \nonumber \\
& \leq \frac{h(\alpha)}{\sqrt{2}}\sqrt{S(\rho || \sigma)}
\end{align}
\end{widetext}
The penultimate inequality is taken from Ref. \citep{ruskai_2014}. In the last step, we apply the Pinsker inequality. Now, taking $\sigma$ as an incoherent state which minimizes the relative entropy of coherence, i.e., the dephased version of $\rho$, the theorem above follows immediately.
\end{proof} Since relative entropy is in turn bounded above by the another popular coherence measure, viz. the $l_1$ norm, a weaker upper bound on the CQA may be given in terms of the $l_1$ norm of coherence as well.

\emph{ A \lowercase{state-dependent quantum Uncertainty Relation}-} Uncertainty relations are one of the fundamental pillars of quantum theory \citep{heisenberg,kennard,robertson}. However, for mixed quantum states, there are two sources of uncertainty present, the first is the classical stochastic randomness in preparing mixed states and the second being the intrinsic quantum randomness. It is thus desirable to separate out these two types of randomness, and especially to formulate uncertainty relations in terms of the quantum part of the uncertainty alone. Luo in his seminal paper \citep{luo} prescribed the Wigner Yanase skew information for quantification of the \emph{quantum part} of uncertainty. However, the uncertainty relations obtained through them have state-dependent lower bounds. Let us recall at this juncture Luo's criteria \citep{luo2,korzekwa} for an entity $Q$ to be a quantifier of the \emph{quantum part} of the uncertainty for an observable $A$.

\begin{enumerate}
\item Q must vanish if the state $\rho$ of the quantum system commutes with the observable $A$.

\item Q must be convex with respect to the state $\rho$
\end{enumerate}

It has already been proved that the CQA follows both those properties when coherence is computed with respect to the eigenbasis of $A$. Indeed, any convex coherence monotone is thus a quantifier of the quantum part of uncertainty, and such resulting uncertainty relations in terms of other popular coherence quantifiers have already been discovered \citep{uttamur}. For a discussion, we refer to \citep{luo2}. Thus, it is imperative that we prove an uncertainty relation in terms of CQA with respect to eigenbases of two (generally non-commuting) observables. Suppose $\lbrace\sigma^Q\rbrace$ are the family of states diagonal in the eigenbasis of a Hermitian operator $Q$. In this case, the following state-dependent uncertainty relation is proved for two observables  $A$ and $B$. 

\noindent \emph{Theorem- If $\sigma^{A}$ and $\sigma^{B}$ are the diagonal states in the respective eigenbases of $A$ and $B$, for which the requisite minimization for the CQA is obtained, then the following uncertainty relation holds}
\beq \sqrt{\left[C_{\alpha} (\rho) \right]_A} + \sqrt{\left[C_{\alpha} (\rho) \right]_B} \geq \frac{1}{2} \sqrt{\alpha (1-\alpha)} ||\ [[\sigma^A, \sigma^B], \rho] \ ||_1   \eeq
\begin{proof}
From the Pinsker inequality, we have
\begin{align} 
& [C_{\alpha} (\rho)]_{A}  \geq  \frac{1}{2} \alpha (1-\alpha) ||[\rho, \sigma^{A}]||_{1}^{2} \nonumber\\
& [C_{\alpha} (\rho)]_B  \geq  \frac{1}{2}\alpha (1-\alpha) ||[\rho, \sigma^{B}]||_{1}^{2}
\label{pinsk}
\end{align}

Now, let us consider the Bianchi identity \beq [[\rho, \sigma^A], \sigma^B] + [[\sigma^A, \sigma^B],\rho] + [[\sigma^B,\rho],\sigma^A] = 0 \eeq and remind ourselves of the Kittaneh's inequality for arbitrary linear operators $A$ and $B$ \beq ||[A,B]|| \leq 2 ||A|| \ ||B|| \eeq

Together with the fact that density matrices have unit norm, they imply
\begin{align}
& ||[[\sigma^A,\sigma^B], \rho]||_1 =  ||[\sigma^A,\rho],\sigma^B] + [[\rho, \sigma^B],\sigma^A] ||_1\nonumber \\
& \leq 2 ||[\rho, \sigma^A] ||_1 + 2 ||[\rho, \sigma^B] ||_1
\end{align}
Now combining this result with \eqref{pinsk} completes the proof.
\end{proof}

It is clear that although the coherence quantifier considered here is entropic in nature, unlike the canonical entropic uncertainty relations \citep{wehner_review}, the lower bound of the above uncertainty relation is state dependent, moreover it is expressed in terms of the commutator between  states diagonal in the eigenbases of the respective observables, thus retaining a flavour of the Robertson form of uncertainty relation, which also has a state dependent lower bound.

\emph{Conclusion and future scope -} To summarize, we have shown that for mixed states, quantum addition, though not in same footing as quantum superposition for pure states, does provide a clear operational significance in quantitatively capturing the quantum coherence. We know that mixing, whether classical, or quantum, increases entropy, but in this letter, we showed that the increase in entropy due to quantum mixing (addition) is always lower than the former. We further showed that the relative entropic difference between quantum and classical mixing is actually related to the coherence content of a mixed state. This agrees with our intuition, since the property of non-commutativity which is responsible for the difference between classical mixture and the qudit addition channel also underlies the theory of quantum coherence. We also derived a state dependent $quantum$ uncertainty relation, which may be considered in conjunction with recent variance-based stronger uncertainty relations \citep{mpur,debasis,mdur,branciardur,workinprogress}, to investigate the link between uncertainty and quantum resources like coherence. From an information theoretic standpoint, the implications of our results for the coherence generating capacity of qudit channels may turn out to be potentially important.

For CV systems or even finite dimensional systems with non-orthogonal bases, the quantification of superposition is trickier. In those systems, the analogs to qudit addition channel considered here, e.g.  the beamsplitter in quantum optical systems, may be useful for resource quantification. Another facet of the present work would be to consider the amount of magic \citep{njp_magic,contextuality_supplies_magic,chiru9} present in the system. If we allow classical convex mixture of two pure stabilizer states, the resultant state is again within the stabilizer polytope. However, performing a qudit addition channel on two stabilizer states may result in a magical output state. The consideration of such resource creation via the qudit additon channels and the related question of calculating the energetic cost of such operations, as well as  mathematical  investigations into the  geometry of quantum states under qudit addition channels should be interesting for future work. 

We acknowledge financial support from the Department of Atomic Energy, Govt. of India. We also thank Lin Zhang for useful feedback.
\bibliography{ur_ref} 
\bibliographystyle{apsrev4-1}
\end{document}